\documentclass[twocolumn]{aastex631}
\shorttitle{Hfs structure of Tm\,{\sc ii}}
\shortauthors{Kebapc{\i} et al.}
\graphicspath{{./}{figures/}}

\usepackage{color}
\usepackage{xcolor}
\usepackage{colortbl}

\newcommand{\sk}[1]{\textcolor{green}{#1}}

\begin{document}

\title{Hyperfine Structure Investigation of Singly Ionized Thulium in FT Spectra}

\author[0000-0001-8885-5545]{Taha Yusuf Kebapc{\i}}
\affiliation{Institute of Graduate Studies in Sciences, Istanbul University,
TR-34452 Beyaz{\i}t, Istanbul, T\"urkiye}
\affiliation{Pamukkale University K{\i}n{\i}kl{\i} Campus,
Denizli Health Services Vocational School of Higher Education,
TR-20070 K{\i}n{\i}kl{\i}, Denizli, T\"urkiye}

\author[0000-0003-2355-4674]{\c{S}eyma Parlatan}
\affiliation{Istinye University, Vocational School of Health Services,
TR-34010, Zeytinburnu, Istanbul, T\"urkiye}

\author[0009-0003-0164-5126]{Sam\.I Sert}
\affiliation{Institute of Graduate Studies in Sciences, Istanbul University,
TR-34452 Beyaz{\i}t, Istanbul, T\"urkiye}

\author[0000-0002-3664-3388]{\.Ipek Kanat \"Ozt\"urk}
\affiliation{Istanbul University, Faculty of Science, Physics Department,
TR-34134 Vezneciler, Istanbul, T\"urkiye}

\author[0000-0002-2428-8163]{G\"on\"ul Ba\c{s}ar}
\affiliation{Istanbul University, Faculty of Science, Physics Department,
TR-34134 Vezneciler, Istanbul, T\"urkiye}

\author[0000-0002-0296-233X]{T\.Imur \c Sah\.In}
\affiliation{Akdeniz University, Faculty of Science, Department of Space Sciences and Technologies, 07058, Antalya, T\"urkiye}

\author[0000-0003-3510-1509]{Sel\c cuk B\.Il\.Ir}
\affiliation{Istanbul University, Faculty of Science, Department of Astronomy and Space Sciences, 34119, Beyaz\i t, Istanbul, T\"urkiye}

\author[0000-0002-6313-5768]{Ruvin Ferber}
\affiliation{University of Latvia, Faculty of Physics, Mathematics and Optometry, 
19 Rainis Bolevard, Riga LV-1586, Latvia}

\author[0000-0002-0710-3470]{Maris Tamanis}
\affiliation{University of Latvia, Faculty of Physics, Mathematics and Optometry, 
19 Rainis Bolevard, Riga LV-1586, Latvia}

\author[0000-0003-4991-9176]{Sophie Kr\"oger}
\affiliation{Hochschule für Technik und Wirtschaft Berlin,
Wilhelminenhofstr. 75A, Berlin D-12459, Germany}



\begin{abstract}

The hyperfine structure of 40 spectral lines of singly ionized thulium (Tm\,{\sc ii}) in emission spectra from a hollow cathode discharge lamp measured with a Fourier transform spectrometer in the wavelength range from 335 nm to 2345 nm has been investigated. As a result of the analysis, the magnetic dipole hyperfine structure constants $A$ for 27 fine structure levels of Tm\,{\sc ii} were determined for the first time. In addition, the values of two magnetic dipole hyperfine structure constants $A$ from the literature were declared incorrect and the corrected values were given.

\end{abstract}



\section{Introduction} \label{sec:intro}

 Rare earth elements, which comprise the lanthanide series in the periodic table of elements, are of significant astrophysical importance for understanding and testing scenarios for the production of heavy elements (e.g., neutron capture elements) in the universe. The study of the spectra of rare-earth elements in laboratory settings is vital in the field of astrophysics, as these elements exhibit numerous atomic and ionic spectral lines in the spectra of the Sun and other more evolved stars \citep{Wan22, Hanke2020, Tia16, Sneden09, Xu03}. The element Tm, with an atomic number of 69, is a rare-earth element whose hyperfine structure is investigated in this study.

The spectra of rare-earth atoms are generally very complex due to the high number of fine structure levels. There are still many gaps in the knowledge of the fine structure and hyperfine structure (hfs) of all lanthanides. Some energetically high-lying configurations of the atoms and especially of the ions have not yet been studied either experimentally or theoretically.

Tm is the thirteenth element in the series of lanthanides. It is one of the few elements to have only one stable isotope. The nuclear spin of the only naturally occurring isotope, $^{169}$Tm, is $I=1/2$. Therefore, only magnetic hfs splitting occurs in Tm spectra, no electric quadrupole hfs. 

In the literature, there are only very few studies on the fine structure and hfs of singly ionized Tm. The study by \citet{Rad20} presents 1099 calculated fine structure energy levels for Tm\,{\sc ii}. Of these 1099 calculated levels, 380 levels are experimentally known \citep{Mar78, Wya11,Kar23}. To the best of our knowledge, up to now, only one article deals with determination of experimental hfs constants of Tm\,{\sc ii} \citep{Man89}. They performed hfs measurements using the technique of collinear fast ion beam laser spectroscopy and experimental magnetic dipole hfs constants $A$ have been determined for eight fine structure energy levels of Tm\,{\sc ii}. Several other publications on Tm\,{\sc ii} deal with the theoretical and experimental investigation of lifetimes and transition probabilities (of Tm\,{\sc ii} and other degrees of ionization of Tm) \citep{Bla94, And96, Wic97, Qui99, Rie99, Xu03, Tia16, Wan22, Rad20}. 

Considering the importance of HFS analysis in astrophysics, it is clear that there are major gaps in the determination of laboratory data. The present work aims to reduce these gaps by determining magnetic dipole hfs constants $A$ for Tm\,{\sc ii}, which have been previously unknown.

{\section{Applications for astrophysics}}
Spectroscopy is one of the most crucial observation techniques in astronomy. By measuring the intensity of radiation of various wavelengths from a star or from an interstellar medium, its chemical composition can be determined, and the temperature of stars can be estimated. The interior of a star emits a continuous black body spectrum. Atoms and ions in the outer part of the star absorb the light emitted by the inner part, resulting in characteristic absorption lines. The relative abundance of particular elements in the star can be estimated from the wavelength, the depth and the shape of an absorption line \citep{Wah02}. The use of advanced telescopes equipped with high-resolution spectrometers allows for the collection of detailed information that is necessary for constructing models of stellar atmospheres. These instruments make it possible to detect even the weakest spectral lines in the spectra obtained. To gain a comprehensive understanding of the synthesis and abundance of elements heavier than iron, one can study the lines of rare-earth elements in the spectra \citep{Mic15}. These types of studies demand an in-depth understanding of the precise configuration of spectral lines, which encompasses hfs and isotope shift.

According to \citet{Cowley98}, the lanthanides are significantly more abundant in HD 101065, also known as Przybylski's star, than in the Sun, with a difference of approximately 10,000 times in their abundance. A similar statement can be made about the lanthanides in the atmospheres of other chemically peculiar stars. Undoubtedly, studies of this nature, which explore the presence of rare-earth elements in stellar atmospheres, emphasize the necessity of employing laboratory-based techniques to examine the spectral lines of these elements. By conducting such characterization studies, it is possible to obtain precise parameters for the stellar atmospheres being studied.

Accurate atomic data are required to analyse stellar spectra, which necessarily involves detailed information on hfs of lines. For many elements, data on hfs in their spectra are scarce or have large measurement uncertainties. This study is intended to take a small step toward developing and extending atomic data that will allow a more detailed analysis of the stellar spectra.

 \subsection{Need for atomic data for Tm\,{\sc ii} from the perspective of stellar spectroscopy}

\begin{figure*}
    \centering
    \includegraphics[width=1.05\linewidth]{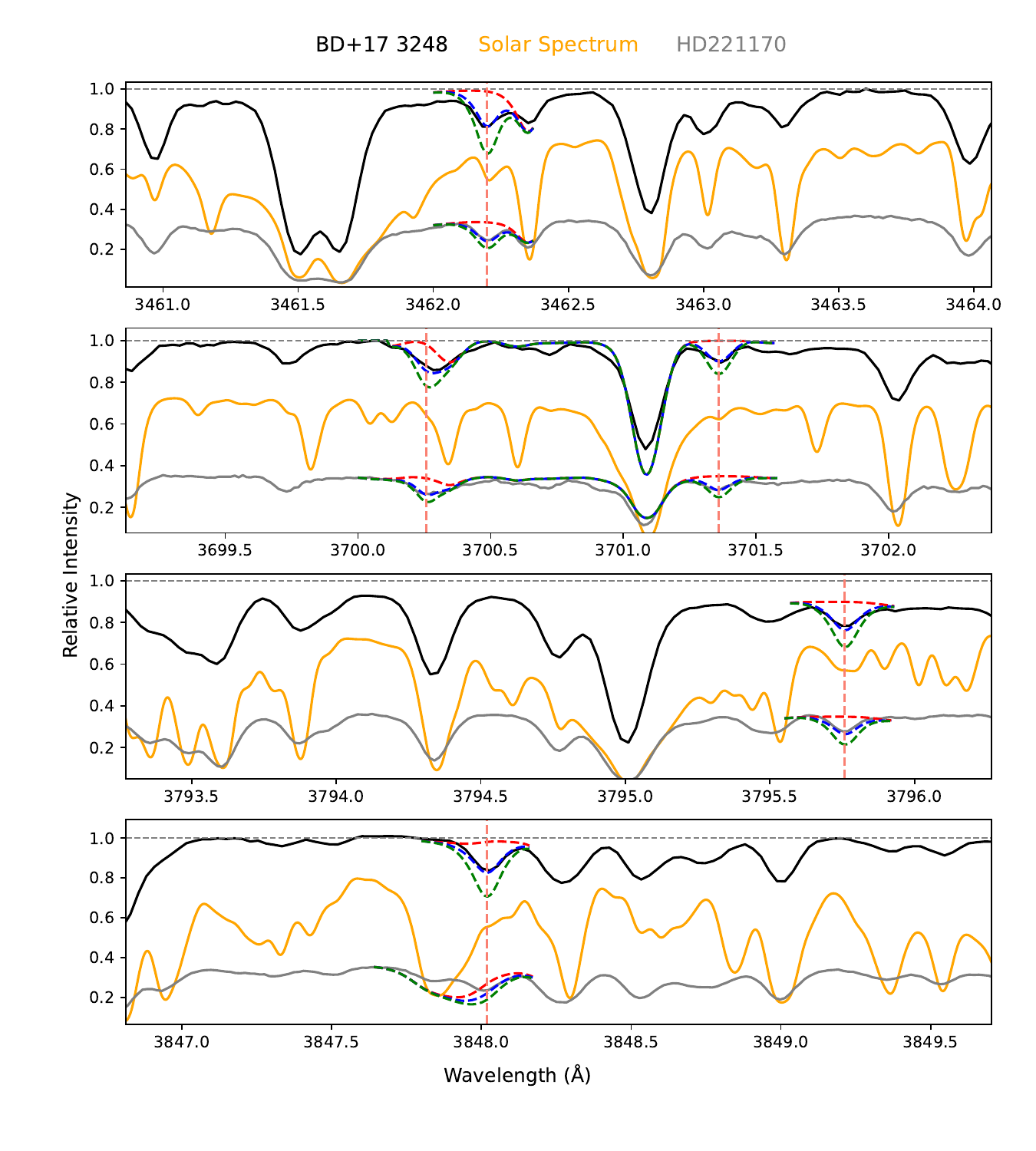}
    \caption{A small region of the Keck I HIRES spectra of BD+17 3248, HD\,221170 and the solar spectrum obtained with Fourier-transform spectrometer and McMath-Pierce telescope. In each panel, the solid lines represent the observed spectrum. The red dotted lines are the spectra computed with no contribution from Tm\,{\sc ii}; the blue dotted line is the best-fitting synthesis; and the green lines are the syntheses computed with Tm abundances altered by +0.5 dex from the best value. The identified Tm\,{\sc ii} lines are indicated with vertical red dashed lines.}
    \label{fig:synth_Tm}
\end{figure*}

\begin{figure*}
    \centering
    \includegraphics[width=0.95\linewidth]{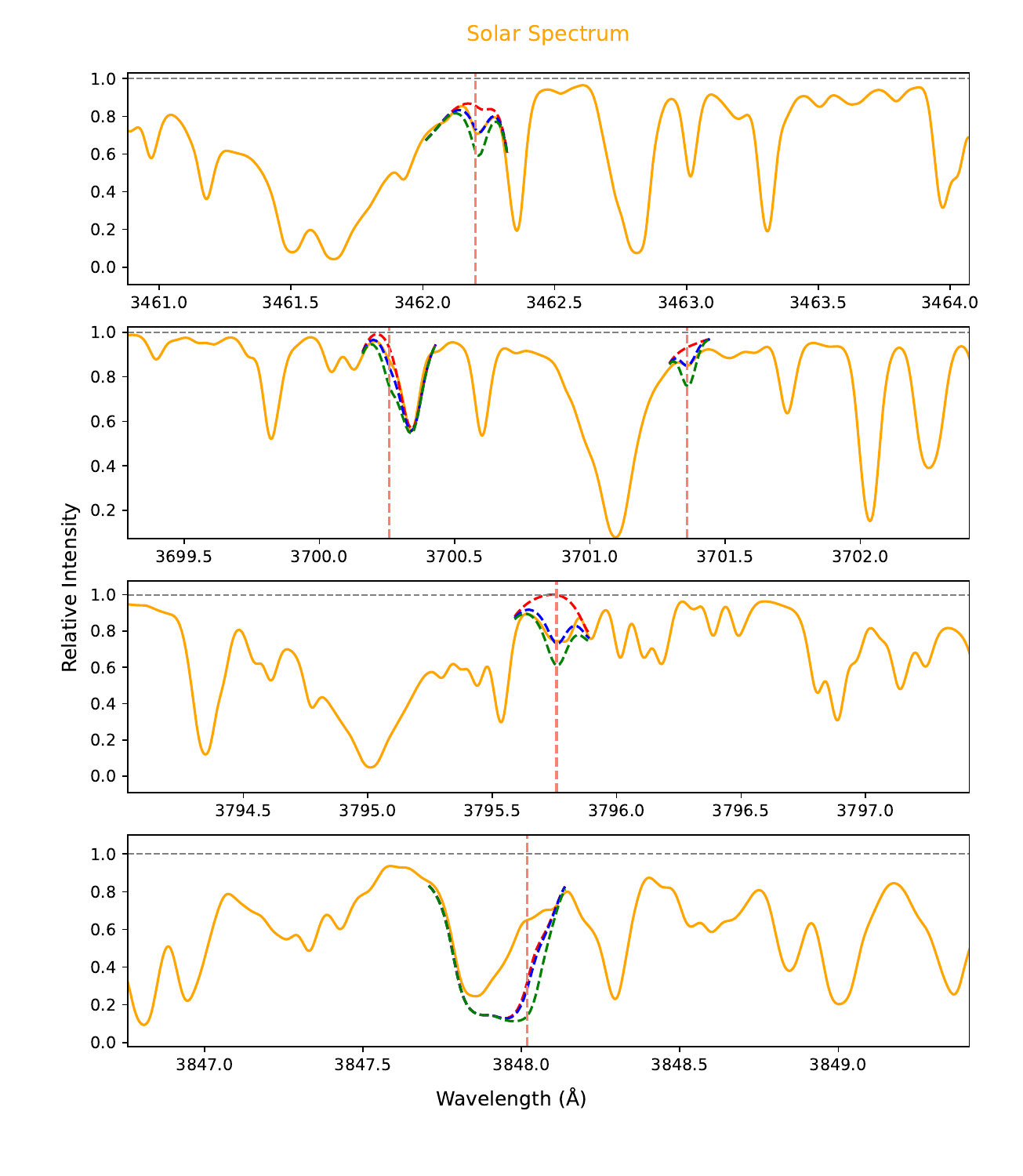}
    \caption{Observed and synthetic spectra of the Sun. The red dotted lines are the spectra computed with no contribution from Tm\,{\sc ii}; the blue line is the best-fitting synthesis; and the green lines are the syntheses computed with Tm abundances altered by +0.5 dex from the best value. The identified Tm\,{\sc ii} lines are indicated with vertical red dashed lines. The solar spectrum is that of \citep{Brault87}.}
    \label{fig:sun_tm}
\end{figure*}

Atomic hfs has an important influence on the determination of stellar element abundances. If not taken into account in spectral analyses, for example, element abundances determined from equivalent widths would be systematically too large and extinction constants would be overestimated \citep{Kurucz93}.
Almost all rare-earth elements detected in stellar spectra are represented by atomic transitions with low excitation potentials ($\leq$ 1 eV), especially in ultraviolet (UV), and are therefore not overly sensitive to stellar effective temperatures. Furthermore, the majority of rare 
earth elements, such as Tm, have not been the subject of systematic analyses of solar and stellar spectra. 
One of the few examples that can be cited as particularly noteworthy is the study conducted by 
\citet{Sneden09}, \citet{Hanke2020} and \citet{Ryabchikova2006}. \citet{Sneden09} reported elemental 
abundances for rare-earth elements, including Tm ($Z$=69), using the high-resolution spectra of a sample 
of solar and metal-poor stars. Spectroscopic analysis by \citet{Hanke2020} for the nuclear benchmark 
star HD\,20 provided an abundance of Tm\,{\sc ii} over four Tm\,{\sc ii} lines at 3700.26 \AA, 3701.36 \AA, 
3795.76 \AA, and 3848.02 \AA. The study by \citet{Ryabchikova2006} revealed three Tm\,{\sc ii} lines for 
the magnetic chemically peculiar star HD\,144897 in the ultraviolet region (3267.39 \AA, 3309.80 \AA, 
and 3462.19 \AA). Tm is one of the least abundant rare-earth elements \citep{Lodders03} and the 
transitions of Tm\,{\sc ii} observed in the solar and other stellar spectra are weak; few can be used 
in abundance analysis. Interesting examples are asymptotic giant branch (AGB) stars, post-AGB stars, and 
barium stars, that is, their photospheric thulium abundances are unknown. A recent study by \citet{Zacs2024} 
reports Tm\,{\sc ii} abundance for HD\,235858 (a post-AGB star) and HD\,204075 (a barium star). It is 
evident that the atomic data for the ionized thulium lines employed in these studies exhibit varying 
degrees of divergence from one line to another. In essence, it is necessary to refine the atomic transition probabilities. In particular, techniques involving combinations of laser-induced fluorescence (for the 
measurement of radiative lifetimes) and Fourier-transform spectroscopy (for the measurement of branching 
fractions) have made it possible to calculate accurate transition probabilities for many elements. For 
example, \citet{Wic97}, in a pioneering study, reported branching fractions for 146 Tm\,{\sc ii} 
lines and laboratory experimental transition probabilities derived from the radiative lifetimes reported 
by \citet{And96}. Figure~\ref{fig:synth_Tm} shows the common Tm\,{\sc ii} transitions in \citet{Sneden09} in the 
spectra of the Sun and the selected metal-poor stars in the near ultraviolet region. High 
resolution Keck I HIRES \citep{Vogt94}\footnote{detailed description at https://koa.ipac.caltech.edu/cgi-bin/KOA/nph-KOAlogin} 
spectra of HD\,221170 ($R=$40\,000 and $S/N=$200) and BD+17\,3248 ($R=$45\,000 and $S/N=$200) were 
provided by Chris Sneden (2024, private communication) and corrected for radial velocity shifts. 
The stars with the same spectra were also studied by \cite{Sneden09} for the elemental 
abundances of rare earth elements. The solar spectrum used is a high-resolution disc-averaged solar 
spectrum obtained with a Fourier-transform spectrometer and McMath-Pierce telescope in the range 3290-
12510 \AA~\citep{Brault87}. 
As seen in Figure~\ref{fig:synth_Tm} and Figure~\ref{fig:sun_tm}, the Tm\,{\sc ii} transitions at 3462.20 \AA, 3700.26 \AA, 3701.36 \AA, 3795.76 \AA, and 3848.02 \AA~are 
visible in the three stellar spectra. The Tm\,{\sc ii} lines at 3700.26 \AA~and 3848.02 \AA~are in a wing of a line in the solar spectrum. The Tm\,{\sc ii} line at 3462.20 \AA~is blended in the three spectra. Similarly, the Tm\,{\sc ii} line at 3700.25 \AA~in the spectrum of BD+17\,3248 and HD\,221170 is also blended and therefore 
unsuitable for element abundance determination. However, the 3701.36 \AA~Tm\,{\sc ii} line exhibits a 
measurable line profile in the spectra of these stars. In order to confirm the existence of thulium lines, spectral regions containing these lines were synthesized. We computed the synthetic spectra with the current version of LTE, one-dimensional (1D) line analysis code MOOG \citep{Sneden1973}. We synthesized multiple synthetic spectra for different thulium abundances which also provides a useful method to identify these lines. To determine the abundance values for the Tm lines, thulium abundance was varied in the models created for the stars using the model atmosphere parameters reported by \cite{Sneden09}. The transition probabilities for these lines were also obtained from \cite{Sneden09}. Details on model atmosphere computation, and synthesis were described in \cite{Sahin2011}, \cite{Sahin2016} and \cite{Sahin2020}. The best fits in the wavelength regions around the Tm\,{\sc ii} lines in the three spectra are shown in Figure~\ref{fig:synth_Tm} and Figure~\ref{fig:sun_tm}. The thulium abundances in the atmosphere of BD+17 \,3248 and HD\,221170 were found to be in excellent agreement with the thulium abundances of the stars reported by \cite{Sneden09}: log$\epsilon$(Tm\,{\sc ii})=-1.13$\pm$0.07 dex for BD+17 3248 and log$\epsilon$(Tm\,{\sc ii})=-1.34$\pm$0.10 dex for HD\,221170. It should be noted that the Tm\,{\sc ii} lines at 3700.26 \AA, 3701.36 \AA, 3795.76 \AA~and 3848.02 
\AA~ are measured by FT spectroscopy in the present work. Their hfs $A$ constants are also determined.
Considering the current state of research, reliable atomic 
data that will enable accurate determination of rare-earth element abundances for the transitions under 
study are as important as obtaining high-resolution and high signal-to-noise ratio stellar spectra.

\section{Experiment}
The FT spectra of the element Tm in the range of 335~nm to 2345~nm were measured at the Laser Center at the University of Latvia. The spectra in the visible (VIS) and near-infrared (IR) spectral range have already been used in earlier work of our group for an analysis of the hfs of atomic Tm \citep{Par22, Keb22}. Detailed information about the experiment can be found in these papers. The essentials are briefly summarized again here.

The Tm plasma was generated in a non-commercial, laboratory-made hollow cathode discharge lamp. A foil of 99.9\% pure Tm was inserted into the cylindrical copper cathode with an inner diameter of 3~mm and a length of 20~mm. The cathode was cooled with liquid nitrogen in order to reduce the Doppler line broadening.

Spectra were measured in the UV, the VIS and the IR wavelength ranges. In the UV and VIS wavelength range a photomultiplier tube (Hamamatsu R928) was used for light detection. In the IR range from 6\,000 to 12\,000~cm$^{-1}$ a room temperature InGaAs diode and in the range from 
4\,000 to 6\,000~cm$^{-1}$ a thermoelectrically cooled TE-InGaAS diode with a Peltier element were used. 

In each wavelength range different spectra were measured with two different buffer gases, argon (Ar) and neon (Ne), as well as two different discharge currents of 50~mA and 70~mA. The buffer gas pressure was one to two mbar. Each spectrum was obtained from an average of at least 50 scans. For all known Tm\,{\sc ii} spectral lines, the signal-to-noise ratio (SNR) was better in Tm-Ar spectra than in the Tm-Ne spectra. The factor SNR(Ar-Tm)/SNR(Ne-Tm) lies between 1.5 and 2.3. 

The wavelength calibration of the spectra was verified by comparing the wavenumbers of atomic and ionic Ar and Ne lines in the recorded FT spectra with the corresponding wavenumbers given in \citep{Wha02,San07} for Ar and \citep{Sal04} for Ne. The absolute positions of the Ar and Ne lines could be determined with an accuracy better than 0.01~cm$^{-1}$ for the entire wavelength region.\\

\section{Hyperfine Structure Analysis}

Due to the nuclear spin $I=1/2$, each fine structure level splits into two hfs levels only. As a consequence, in Tm spectra each fine structure transition consists of three or four individual hfs components. The two components with $\Delta F$ = $\Delta J$ are strong and carry almost the entire intensity of the line. The one or two other components are weak (see for example \citep{Cow81}). Our FT spectra display only a limited number of lines featuring two distinct, clearly resolved hyperfine components. Notwithstanding the presence of such lines, the weak components are often positioned within the Doppler width of the strong components, thereby posing a challenge to their precise localization. When the weak components are outside the Doppler width of the strong components, they often disappear in the noise. Additionally, some strong lines show slight asymmetry, which makes it even more difficult to locate the small components. For these reasons, the analysis of the hfs of Tm in our spectra is particularly difficult. This is true for both Tm atoms and Tm ions.

\begin{table}
\caption{Experimental magnetic dipole hfs constants $A$ determined by \citet{Man89}.}
\label{tab:t1}
\begin{tabular}{r@{}l c c r@{\,}l}
\hline
\multicolumn{2}{c}{$E$ in cm$^{-1}$} &
$J$ & $p$ &
\multicolumn{2}{c}{$A$ in MHz} \\[1mm]
\hline
 8\,769&.68 & 2 & o &   914.9&(1.0)*  \\
 8\,957&.47 & 3 & o & -1537.7&(0.8)   \\
17\,624&.65 & 2 & o &  -341.8&(1.3)   \\
21\,713&.74 & 3 & o &  -313.5&(1.9)   \\
25\,980&.02 & 3 & e &   -89.8&(0.7)*  \\
26\,578&.77 & 3 & e &  -136.6&(1.0)   \\
34\,913&.84 & 3 & e &  -493.2&(1.6)   \\
39\,162&.07 & 4 & e &  -614.9&(0.9)   \\
\hline
\end{tabular}
\mbox{}\\[-0.5mm]

$p$: parity (o: odd, e: even)\\
\mbox{}* value proven incorrect in this work; for corrected value see Table~\ref{tab:t3}
\end{table}

In a transition, two energy levels are involved, both with their respective magnetic dipole hfs constant $A$. If the $A$ constants of both levels are unknown and if the position of the weak components cannot be clearly determined, it is not possible to derive the values of both $A$ constants from only one experimental wave number separation between line components, i.e.\ wave number difference between the two strong components. Or, in other words, to determine a value for a new hfs constant $A$ from the separation between the two strong components, the hfs constant $A$ of the other level involved in the transition has to be known.   

As discussed above, hfs constants $A$ of Tm~II are known for only eight levels up to now \citep{Man89}. These values are listed in Table~\ref{tab:t1}. With this in mind, spectral lines were selected for our study, which show two well-resolved strong hyperfine components. In the first step, a set of lines was compiled that are connected to the levels with hfs $A$ constants known from literature, which are listed in Table~\ref{tab:t1}.
These lines are hereafter called \textcolor{violet}{A}-lines. 
However, only six of the eight levels in Table 1 were included in the study.
For the remaining two energetically high-lying levels no suitable spectral transitions were available in our spectra.

\begin{figure*}[h] 
\plotone{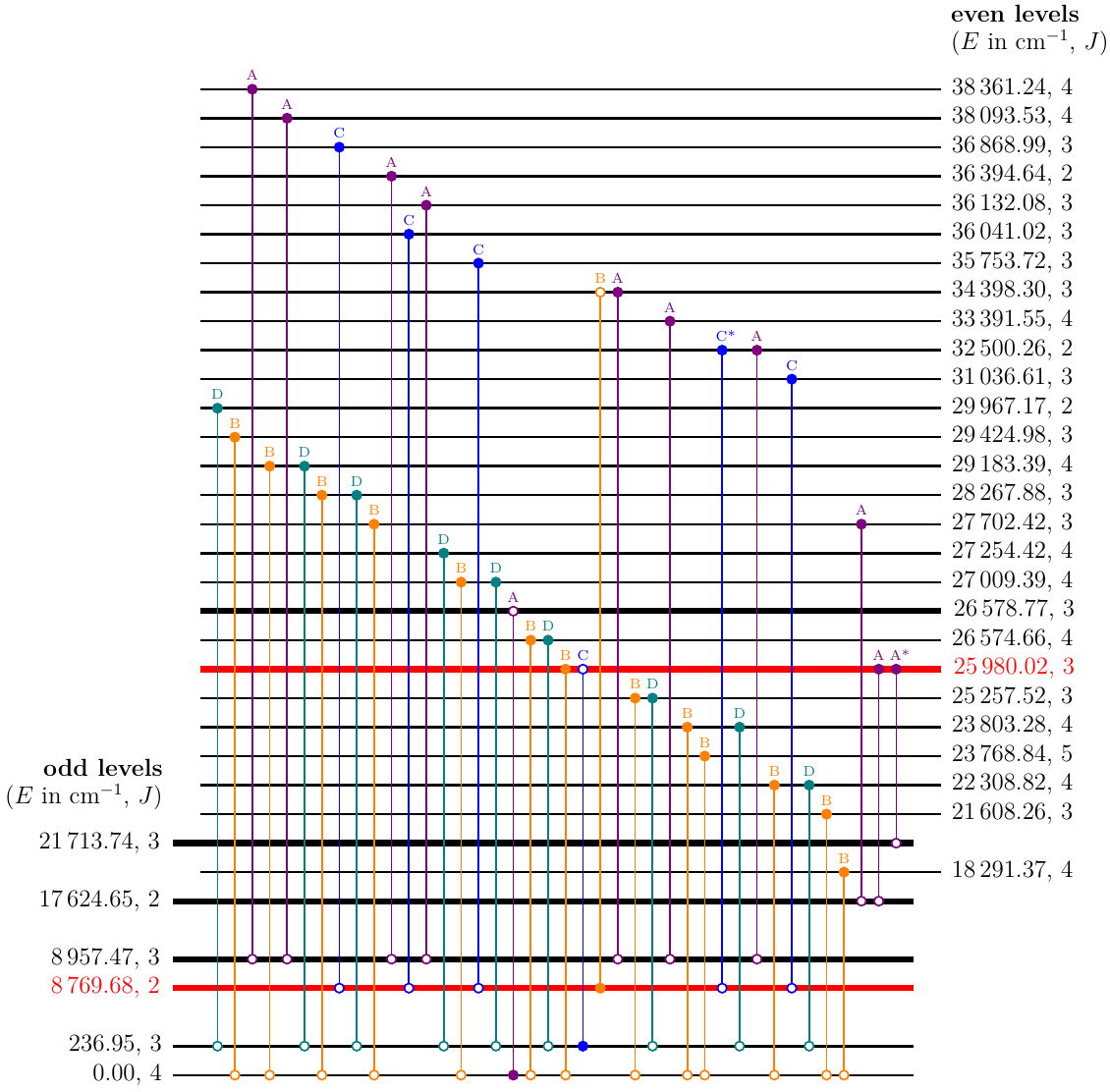}
\caption{Excerpt from the Tm\,{\sc ii} term scheme with all involved energy levels and the corresponding transitions analysed in this work; horizontal lines correspond to an energy level, vertical lines to a spectral line; the energy levels with thick horizontal lines are those with known $A$-values from the literature \citep{Man89}; the energy levels in red are those the $A$ constants of which were proven incorrect in this work; the order of the lines (from left to right) corresponds to the order of the lines in Table~\ref{tab:t2}, i.e. the lines are sorted by ascending wavelength;
the different types of lines are indicated by the letters A  to C above the anchor point on the upper level. For better clarity, the colouring has also been adapted to the letters. A hollow circle at the start or end point of a line means that for this line the hfs $A$ constant of this level was fixed during the fit. Accordingly, a full circle at the start or end point of a line means that for this line the hfs $A$ constant of this level was adjusted as a free parameter in the fit. 
 \label{fig:f1}}
\end{figure*}

In the second step the lines were chosen that were connected to the levels with $A$ constants known from the previous step (hereafter referred to as \textcolor{orange}{B}-lines). This step was repeated two more times yielding the \textcolor{blue}{C}-lines and \textcolor{teal}{D}-lines. 
The rating/assignment of the lines as \textcolor{violet}{A}-, {B}-, \textcolor{blue}{C}- or \textcolor{teal}{D}-lines was done to clearly mark whether the lines are directly connected to a level with an $A$ constant known from the literature or whether they were connected to the system in later steps. The colouring only serves to emphasise this even more clearly, especially in Figure~\ref{fig:f1}, but is also used in the tables. 
 
The program Element \citep{Win03, Win16} was used to select and classify the lines.
In addition to the lines analysed with two resolved peaks, three lines with unresolved strong components have been added to the analysis to complete the system of lines.
These lines were added in order to check the consistency of the hfs data. Lines with unresolved strong hfs components were only included in the investigation if there were no other classification suggestions for them and thus the assignment was unambiguous.
In total, a set of 40 lines was used for the results presented here. 

There are more Tm~II lines observed in the FT spectrum, but they either have unresolved strong hfs components or none of the levels involved has a known hfs $A$ constant, or there are several classification proposals, so that a clear assignment is not possible. Therefore, we have limited our investigation to these 40 lines.

When the lines in the FT spectrum were analysed with the magnetic dipole hfs constants given in \citet{Man89}, some incompatibilities were observed. The simulated intensity ratio and the position of the hfs components showed no agreement with the experimental structure. These problems could be solved by declaring two of the values for the hfs $A$ constants given by \citet{Man89} to be incorrect.
We have refrained here from explaining in detail the individual steps and aberrations on the way to this result. 
After presenting our results in this Section, in the next Section 
we have put together a few brief reflections on the hfs $A$ values from \citet{Man89} declared incorrect, which we would like to use to reinforce our assumption.

\begin{table*}
\caption{Tm~II lines measured via FT spectroscopy and analysed to determine new hyperfine structure constants $A$. The levels printed in bold are those whose $A$ constant has been determined with the respective line.}
\label{tab:t2}
\hspace*{-10mm}
\begin{tabular}{r@{}l r@{}l r c  l | r@{\,}l  c | r@{\,}l  c}
\hline \hline
\multicolumn{7}{c|}{line} &
\multicolumn{3}{c|}{upper level} &
\multicolumn{3}{c}{lower level} \\
\hline
\multicolumn{2}{c}{$\lambda_{\mathrm{air}}$} & 
\multicolumn{2}{c}{$\sigma_{\mathrm{obs}}$} &
\multicolumn{1}{c}{SNR} &
\multicolumn{1}{c}{com.} &
\multicolumn{1}{c|}{ass.} &
\multicolumn{2}{c}{$E$} &
$J$ &
\multicolumn{2}{c}{$E$} &
$J$ \\
\multicolumn{2}{c}{in nm} & 
\multicolumn{2}{c}{in cm$^{-1}$} &
\multicolumn{1}{c}{} &
\multicolumn{1}{c}{} &
\multicolumn{1}{c|}{} &
\multicolumn{2}{c}{in cm$^{-1}$} &
&
\multicolumn{2}{c}{in cm$^{-1}$} &
\\[1mm]
\hline
   336&.261 & 29\,730&.23 & 240  &     & \textcolor{teal}{D}       & \textbf{29\,967}&\textbf{.17} & 2 &              236&.95 & 3          \\
   339&.750 & 29\,424&.98 & 80   &     & \textcolor{orange}{B}     & \textbf{29\,424}&\textbf{.98} & 3 &                0&.00 & 4          \\
   339&.995 & 29\,403&.78 & 50   &     & \textcolor{violet}{A}     & \textbf{38\,361}&\textbf{.24} & 4 &           8\,957&.47 & 3          \\
   342&.562 & 29\,183&.40 & 50   &     & \textcolor{orange}{B}     & \textbf{29\,183}&\textbf{.39} & 4 &                0&.00 & 4          \\
   343&.119 & 29\,136&.07 & 37   &     & \textcolor{violet}{A}     & \textbf{38\,093}&\textbf{.53} & 4 &           8\,957&.47 & 3          \\
   345&.367 & 28\,946&.45 & 245  &     & \textcolor{teal}{D}       & \textbf{29\,183}&\textbf{.39} & 4 &              236&.95 & 3          \\
   353&.657 & 28\,267&.88 & 125  &     & \textcolor{orange}{B}     & \textbf{28\,267}&\textbf{.88} & 3 &                0&.00 & 4          \\
   355&.779 & 28\,099&.31 & 83   &     & \textcolor{blue}{C}       & \textbf{36\,868}&\textbf{.99} & 3 &           8\,769&.68 & 2          \\
   356&.647 & 28\,030&.94 & 140  &     & \textcolor{teal}{D}       & \textbf{28\,267}&\textbf{.88} & 3 &              236&.95 & 3          \\
   360&.876 & 27\,702&.44 & 385  &     & \textcolor{orange}{B}     & \textbf{27\,702}&\textbf{.42} & 3 &                0&.00 & 4          \\
   364&.365 & 27\,437&.16 & 150  &     & \textcolor{violet}{A}     & \textbf{36\,394}&\textbf{.64} & 2 &           8\,957&.47 & 3          \\
   366&.581 & 27\,271&.34 & 95   &     & \textcolor{blue}{C}       & \textbf{36\,041}&\textbf{.02} & 3 &           8\,769&.68 & 2          \\
   367&.886 & 27\,174&.59 & 65   &     & \textcolor{violet}{A}     & \textbf{36\,132}&\textbf{.08} & 3 &           8\,957&.47 & 3          \\
   370&.026 & 27\,017&.48 & 680  &     & \textcolor{teal}{D}       & \textbf{27\,254}&\textbf{.42} & 4 &              236&.95 & 3          \\
   370&.136 & 27\,009&.40 & 785  & \#  & \textcolor{orange}{B}     & \textbf{27\,009}&\textbf{.39} & 4 &                0&.00 & 4          \\
   370&.484 & 26\,984&.03 & 75   &     & \textcolor{blue}{C}       & \textbf{35\,753}&\textbf{.72} & 3 &           8\,769&.68 & 2          \\
   373&.412 & 26\,772&.46 & 310  &     & \textcolor{teal}{D}       & \textbf{27\,009}&\textbf{.39} & 4 &              236&.95 & 3          \\
   376&.133 & 26\,578&.78 & 510  &     & \textcolor{violet}{A}     &          26\,578&.77          & 3 &       \textbf{0}&\textbf{.00} & 4 \\
   376&.191 & 26\,574&.67 & 305  &     & \textcolor{orange}{B}     & \textbf{26\,574}&\textbf{.66} & 4 &                0&.00 & 4          \\
   379&.576 & 26\,337&.72 & 190  &     & \textcolor{teal}{D}       & \textbf{26\,574}&\textbf{.66} & 4 &              236&.95 & 3          \\
   384&.802 & 25\,980&.03 & 140  &     & \textcolor{orange}{B}     & \textbf{25\,980}&\textbf{.02} & 3 &                0&.00 & 4          \\
   388&.344 & 25\,743&.09 & 95   &     & \textcolor{blue}{C}       &          25\,980&.02          & 3 &     \textbf{236}&\textbf{.95} & 3 \\
   390&.078 & 25\,628&.62 & 80   &     & \textcolor{orange}{B}     &          34\,398&.30          & 3 &  \textbf{8\,769}&\textbf{.68} & 2 \\
   392&.958 & 25\,440&.83 & 90   &     & \textcolor{violet}{A}     & \textbf{34\,398}&\textbf{.30} & 3 &           8\,957&.47 & 3          \\
   395&.809 & 25\,257&.53 & 455  &     & \textcolor{orange}{B}     & \textbf{25\,257}&\textbf{.52} & 3 &                0&.00 & 4          \\
   399&.558 & 25\,020&.58 & 115  &     & \textcolor{teal}{D}       & \textbf{25\,257}&\textbf{.52} & 3 &              236&.95 & 3          \\
   409&.149 & 24\,434&.09 & 8    &     & \textcolor{violet}{A}     & \textbf{33\,391}&\textbf{.55} & 4 &           8\,957&.47 & 3          \\
   419&.992 & 23\,803&.29 & 280  &     & \textcolor{orange}{B}     & \textbf{23\,803}&\textbf{.28} & 4 &                0&.00 & 4          \\
   420&.600 & 23\,768&.85 & 130  &     & \textcolor{orange}{B}     & \textbf{23\,768}&\textbf{.84} & 5 &                0&.00 & 4          \\
   421&.278 & 23\,730&.59 & 8    &     & \textcolor{blue}{C}$^*$   & \textbf{32\,500}&\textbf{.26} & 2 &           8\,769&.68 & 2          \\
   424&.215 & 23\,566&.34 & 2085 & \#  & \textcolor{teal}{D}       & \textbf{23\,803}&\textbf{.28} & 4 &              236&.95 & 3          \\
   424&.639 & 23\,542&.80 & 14   &     & \textcolor{violet}{A}     & \textbf{32\,500}&\textbf{.26} & 2 &           8\,957&.47 & 3          \\
   448&.127 & 22\,308&.84 & 860  & \#  & \textcolor{orange}{B}     & \textbf{22\,308}&\textbf{.82} & 4 &                0&.00 & 4          \\
   448&.970 & 22\,266&.94 & 32   &     & \textcolor{blue}{C}       & \textbf{31\,036}&\textbf{.61} & 3 &           8\,769&.68 & 2          \\
   452&.938 & 22\,071&.89 & 140  &     & \textcolor{teal}{D}       & \textbf{22\,308}&\textbf{.82} & 4 &              236&.95 & 3          \\
   462&.656 & 21\,608&.26 & 85   &     & \textcolor{orange}{B}     & \textbf{21\,608}&\textbf{.26} & 3 &                0&.00 & 4          \\
   546&.554 & 18\,291&.38 & 38   &     & \textcolor{orange}{B}     & \textbf{18\,291}&\textbf{.37} & 4 &                0&.00 & 4          \\
   992&.010 & 10\,077&.78 & 45   &     & \textcolor{violet}{A}     & \textbf{27\,702}&\textbf{.42} & 3 &          17\,624&.65 & 2          \\
1\,196&.507 &  8\,355&.37 & 180  &     & \textcolor{violet}{A}     & \textbf{25\,980}&\textbf{.02} & 3 &          17\,624&.65 & 2          \\
2\,343&.319 &  4\,266&.29 & 10   &     & \textcolor{violet}{A}$^*$ & \textbf{25\,980}&\textbf{.02} & 3 &          21\,713&.74 & 3          \\
\hline
\end{tabular}
\tablecomments{
$\sigma_{\mathrm{obs}}$ is the observed wavenumber; uncertainties of these values are $< 0.015$~cm$^{-1}$; 
SNR: signal to noise ratio in the Tm-Ar spectra measured with a discharge current of 70~mA;
level energy $E$ according to \citet{Mar78,Kar23};
com.: comment; {\#}: lines with unresolved strong components;
ass.: assignment, see text for explanation; *: weak and noisy lines not taken into account when calculating the mean values of $A$ constants; 
all upper levels have even parity, all lower levels have odd parity.}
\end{table*}

In Table~\ref{tab:t2} the list of all lines used for determination of the hfs constants $A$ is given. 
The energy values in the sixth and eighth columns are all from \citet{Mar78,Kar23}.
The observed wavenumbers $\sigma_{\mathrm{{obs}}}$ listed in the second column are the centre of gravity of the hfs that results from the fits of our FT spectra.
The mean value from the four measurements is given. 
The standard deviation for the wavenumber in the four measurements is dependent on the SNR and FWHM, but in all cases it is less than 0.005~cm$^{-1}$, for most lines even less than 0.001~cm$^{-1}$.
Together with the uncertainty from the calibration, this results in an overall uncertainty of less than 0.015~cm$^{-1}$ for the wavenumbers. 

The air wavelengths are calculated from the wavenumber values according to the five-parameter formula of \citet{Pec72}. In the third column the SNR in our Tm-Ar spectrum is given.

 Figure~\ref{fig:f1} shows the term scheme with all lines listed in Table~\ref{tab:t2}. The information in Table~\ref{tab:t2} and Figure~\ref{fig:f1} is partly redundant, but this is for the sake of clarity.

The level at 25980.02~cm$^{-1}$ participates in two \textcolor{violet}{A}-lines, but one of them is very weak and noisy. 
In our term scheme the low-lying level at 236.95~cm$^{-1}$ is attached to the level at 25980.02~cm$^{-1}$.
This is why 
the hfs $A$ constant of the level at 236.95~cm$^{-1}$, as well as all evaluations based in the next step on this level depend on the quality of the determination of the hfs $A$ constant of the level at 25980.02~cm$^{-1}$. 
Therefore, the result of an additional \textcolor{orange}{B}-line is taken into 
account for the determination of the $A$ constant of the level at 25980.02~cm$^{-1}$
before 
the value for the 236.95~cm$^{-1}$ level was determined.
The transition from 25980.02~cm$^{-1}$ to 236.95~cm$^{-1}$ was therefore classified as a \textcolor{blue}{C}-line (and not as a \textcolor{orange}{B}-line). Consequently, all lines connected to this level at 236.95~cm$^{-1}$ were rated as \textcolor{teal}{D}-lines. 
\medskip

For all selected lines an appropriate portion of about 0.5~cm$^{-1}$ width has been extracted for analysis from the entire FT spectrum. 
This has been done for all four measured spectra, i.e. the Tm-Ar and Tm-Ne spectra in each case measured with a current of 50~mA or 70~mA.
The extracted portions of experimental spectra of the lines then were fitted using the computer program Fitter \citep{Zei22}. Voigt profile functions for each hfs component with intensity ratios corresponding to the theoretical intensities have been applied (see for example \citep{Cow81}). The fitting parameters have been the hfs constants $A$ of one of the energy levels, the centre of gravity of the line, the overall intensity of the line and the full width at half maximum (FWHM) of the Gaussian and Lorentzian component of the Voigt profile. 

\begin{figure}[h] 
\includegraphics[width=0.45\textwidth]{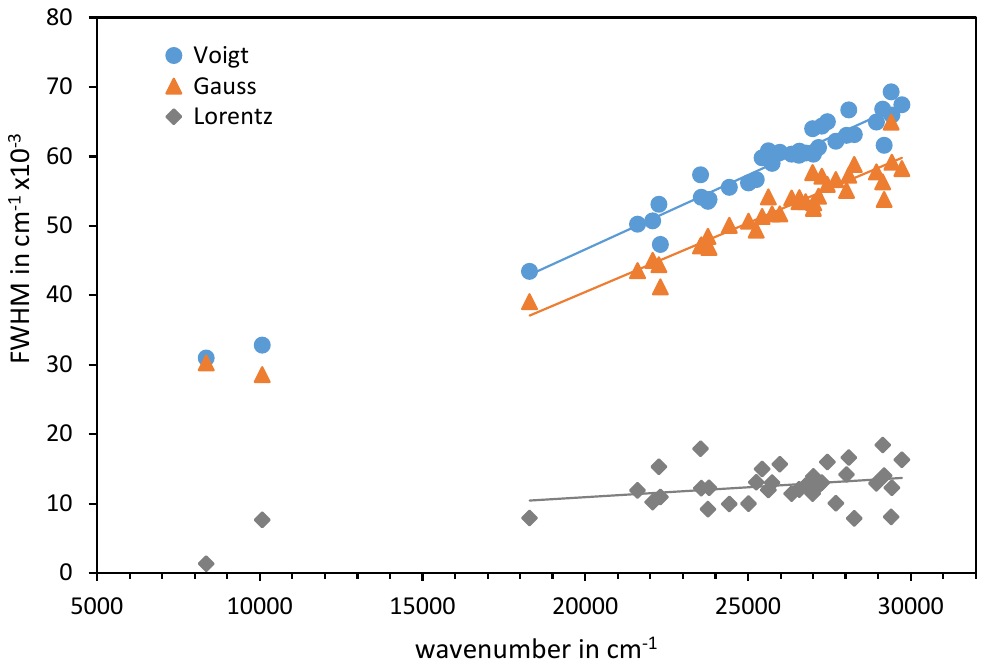}
\caption{
FWHM as a function of the transition wavenumber for the Voigt profile as well as for the Lorentzian and the Gaussian parts that resulted from the fit of the hyperfine structure, fitted in each case by straight lines. All investigated lines in the VIS are recorded except those for which the FWHM had been fixed during the fit.  
}
\label{fig:fx_fwhm}
\end{figure}

Figure~\ref{fig:fx_fwhm} shows the FWHM values as a function of wavenumber for the Lorentzian and Gaussian parts of the Voigt profile function as well as the FWHM values of the Voigt profile. The values were derived from the hyperfine structure fit. 
In Figure~\ref{fig:fx_fwhm} a linear fit is inserted into the diagram for each FWHM data set of all lines in the VIS spectral range. In the IR range, the experimental conditions are slightly different and the two lines in the IR were not taken into account for the linear fit. The slope of the Voigt FWHM is almost identical to the slope of the FWHM of the Gaussian part.
As in previous work of our group reporting on measurements with the FT spectrometer on other elements \citep{Kro10,Guz14}, it can be seen that the FWHM of the Lorentzian part is almost independent of the wavelength.
The Lorentzian component primarily represents the instrumental function of the FT spectrometer, which was around 10$\cdot 10^{-3}$~cm$^{-1}$. The natural linewidth for the analysed lines is well below $10\cdot 10^{-3}$~cm$^{-1}$ and its contribution is negligible. Due to the Doppler broadening in the hollow cathode lamp, the Gaussian component of the linewidth varies between $30\cdot 10^{-3}$~cm$^{-1}$ for small wavenumbers and $60\cdot 10^{-3}$~cm$^{-1}$ for large wavenumbers.

In the fit of the three lines with unresolved strong components, the Gaussian and Lorentzian components of the line width were not fitted but fixed at values determined according to the linear trends shown in Figure~\ref{fig:fx_fwhm}.

\begin{figure*}
\gridline{\fig{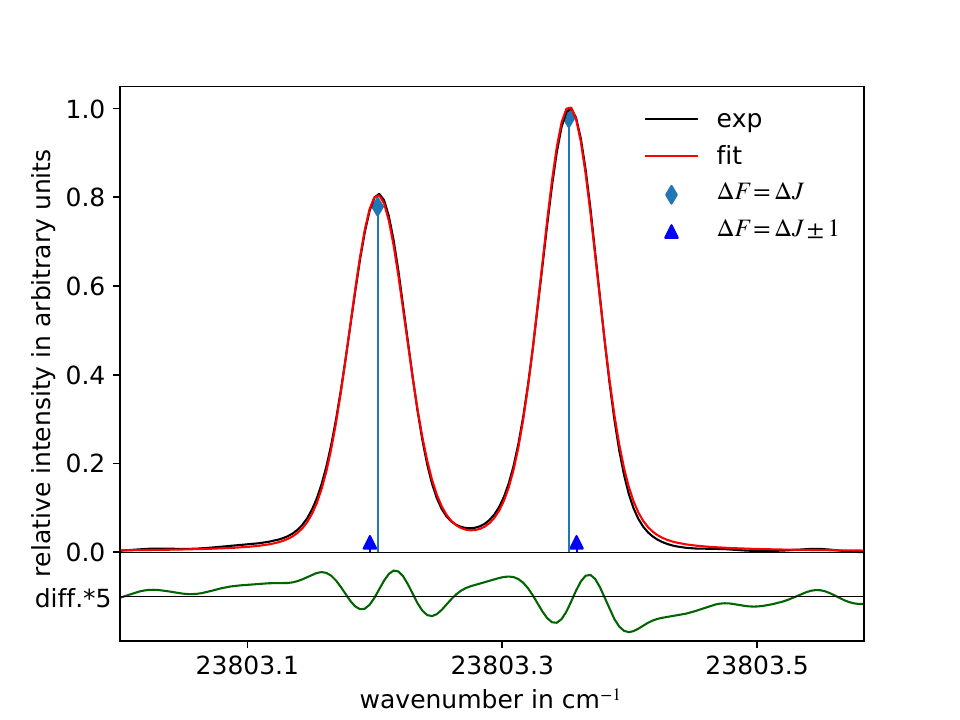}{0.4\textwidth}{(a)}
          \fig{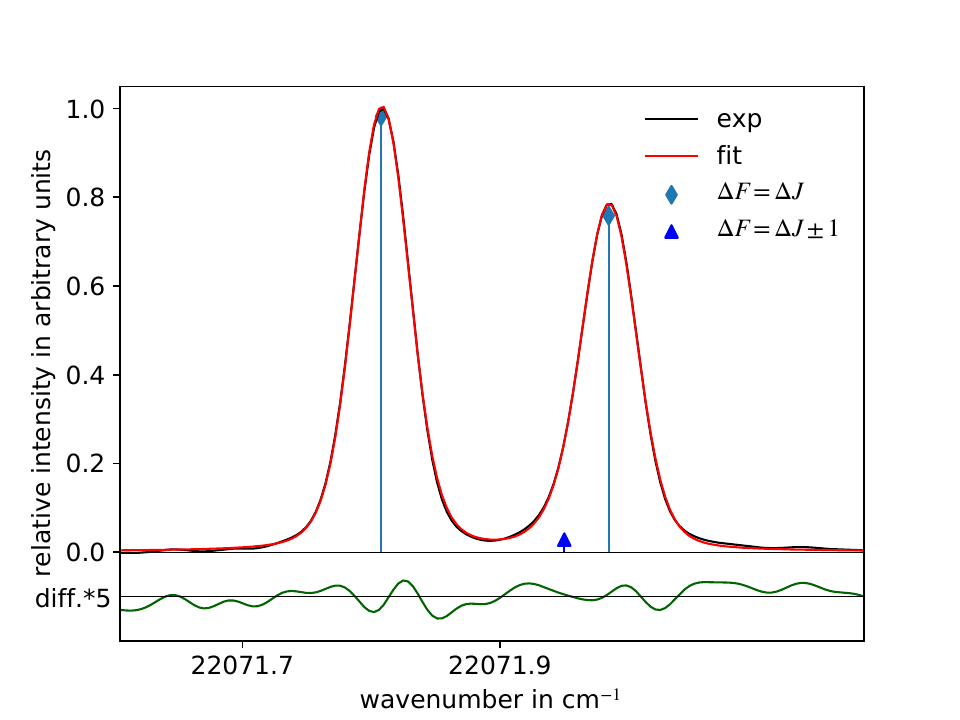}{0.4\textwidth}{(b)}
					}
\gridline{\fig{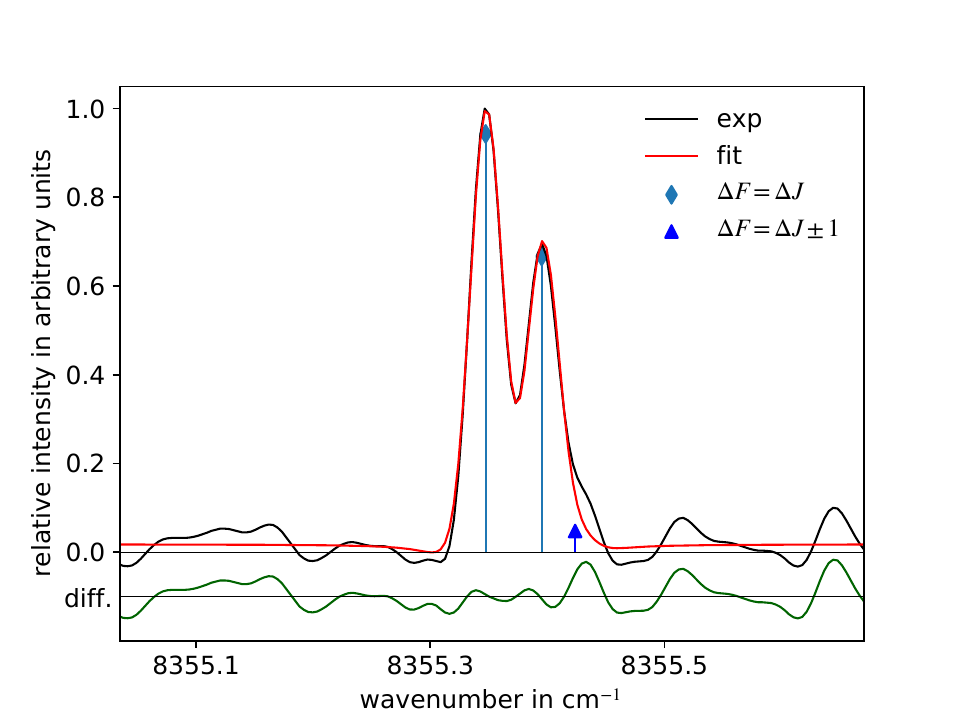}{0.4\textwidth}{(c)}
          \fig{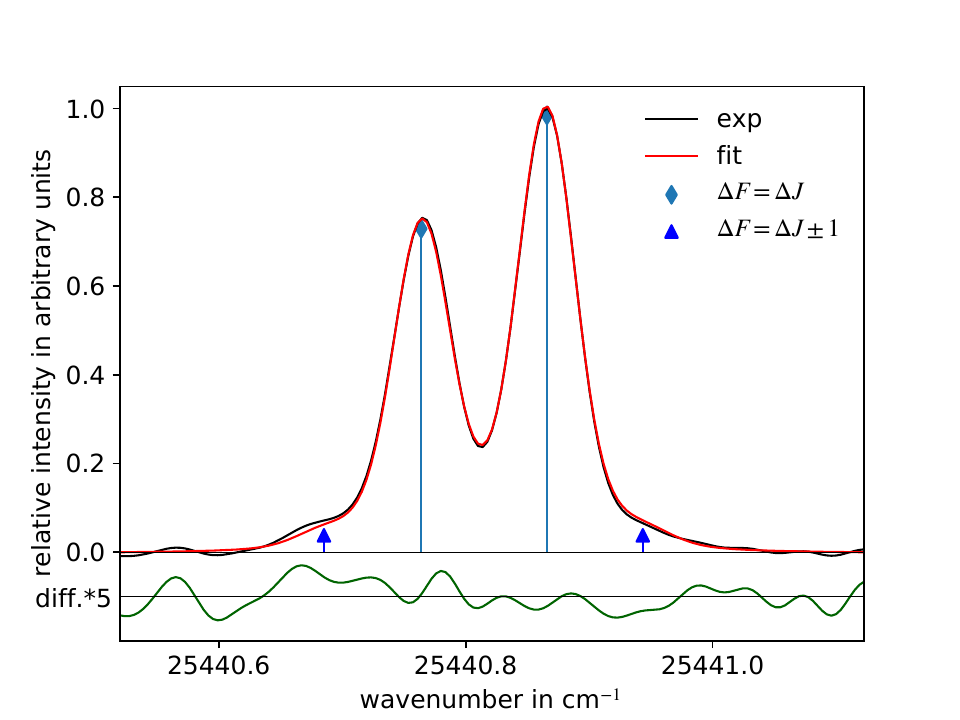}{0.4\textwidth}{(d)}   
          }
\gridline{\fig{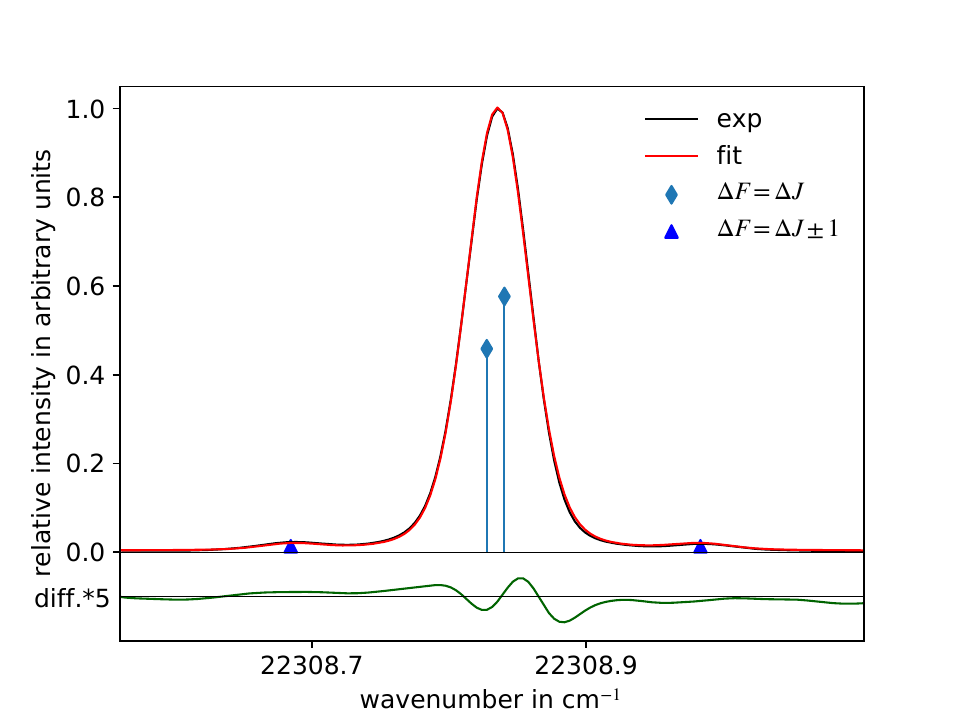}{0.4\textwidth}{(e)}
					\fig{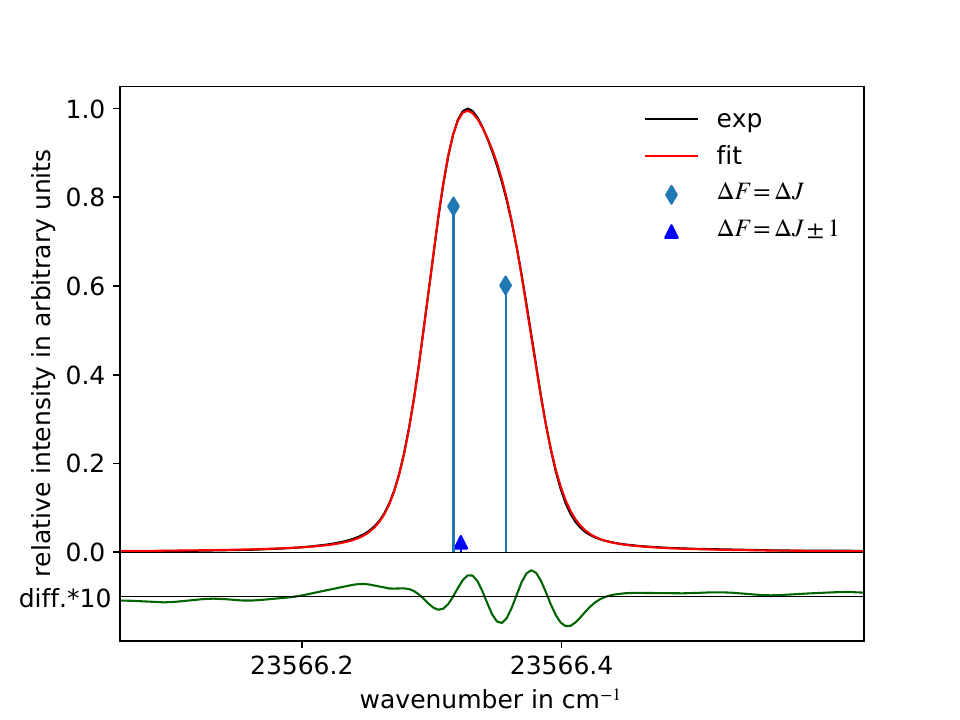}{0.4\textwidth}{(f)}
          }
\caption{Examples of experimental spectra with a corresponding best fit. In the lower part, the residual (diff.) is given. All the experimental spectra are from a Tm-Ar hollow cathode with a discharge current of 70 mA. \\
a) transition 23\,803.28~cm$^{-1}$ ($J = 4$) to       0.00~cm$^{-1}$ ($J = 4$) at $\sigma_{\mathrm{obs}} = 23\,803.29$~cm$^{-1}$,\\ 	
b) transition 22\,308.82~cm$^{-1}$ ($J = 4$) to     236.95~cm$^{-1}$ ($J = 3$) at $\sigma_{\mathrm{obs}} = 22\,071.89$~cm$^{-1}$,\\ 	
c) transition 25\,980.02~cm$^{-1}$ ($J = 3$) to 17\,624.65~cm$^{-1}$ ($J = 2$) at $\sigma_{\mathrm{obs}} =  8\,355.37$~cm$^{-1}$,\\	
d) transition 34\,398.30~cm$^{-1}$ ($J = 3$) to  8\,957.47~cm$^{-1}$ ($J = 3$) at $\sigma_{\mathrm{obs}} = 25\,440.83$~cm$^{-1}$,\\ 
e) transition 22\,308.82~cm$^{-1}$ ($J = 4$) to       0.00~cm$^{-1}$ ($J = 4$) at $\sigma_{\mathrm{obs}} = 22\,308.84$~cm$^{-1}$,\\
f) transition 23\,803.28~cm$^{-1}$ ($J = 4$) to     236.95~cm$^{-1}$ ($J = 3$) at $\sigma_{\mathrm{obs}} = 23\,566.34$~cm$^{-1}$. 		
\label{fig:f2}}
\end{figure*}

Some examples of the hfs spectra are shown in Figure~\ref{fig:f2}. 
Panels (a) and (b) show cases where the weak components are not visible in the experimental spectrum because they lie within the Doppler width of the strong components; one example is with $\Delta J = 0$ and another one with $\Delta J = 1$.
Panel (c) shows the case where the weak component is smaller than the noise level. 
Panel (d) shows one of the few cases where the weak components are barely visible in the experimental spectrum. 
The last two examples concern the lines with unresolved strong components but with either faintly recognisable weak components, see (e), or with a significant asymmetry allowing a clear positioning of the two strong hfs components, see (f). 

In the lower part of Figures~\ref{fig:f2} from (a) to (f), the residual is plotted, i.e. the difference between the experimental and the best-fitted curves.
In order to better illustrate the deviations, the difference is multiplied by a suitable factor. It should be noted that the factor is different for the different lines. 
Three of the lines, (b), (c) and (d), that are moderately strong or weak show an even distribution of noise. 
The three other lines, (a), (e) and (f) have much higher intensities (see the third column of Table~\ref{tab:t2}).  
For these lines, a slight asymmetry in the profiles is clearly visible. 
The asymmetry reduces the quality of the fit, in particular for lines with high SNR. We are not able to assign this asymmetry to a specific cause.
The asymmetry cannot be taken into account with the Fitter program, however, its influence on the determination of the position of the strong hfs components is significantly smaller than the given uncertainty of the $A$ constants.

The resulting magnetic dipole hfs constants $A_{\mathrm{exp}}$ are listed in Table~\ref{tab:t3}. 
$A_{\mathrm{exp}}$ are the mean values from the four measurements (Tm-Ar and Tm-Ne measured with the 70~mA discharge current and Tm-Ar and Tm-Ne measured with the 50~mA discharge current).
In the fifth column, the observed wavenumber of the line used for determination is listed.
In nine cases, the hfs $A$ constants could be determined with two different transitions, in one case even with three lines. In these cases, the question arises whether the $A$ constant of the lower or the upper level should be adjusted with the second line. 
In all cases the upper level was selected for adjustment, because of the hfs $A$ constants already determined in the first step the value of the lower level had a smaller uncertainty. 
The hfs $A$ constants determined from two lines all agree with each other within the uncertainty limits.  

When calculating the uncertainties of the hfs constants $A_{\mathrm{exp}}$ listed in Table~\ref{tab:t3}, the purely statistical approach often yields a too-low value for the uncertainty, as the statistical approach does not take into account the approximations and limitations inherent in the model. Therefore, using only the standard deviation of the mean based on a sample $s_A$ does not correctly reflect the uncertainty of the $A_{\mathrm{exp}}$ values. 
In order to obtain a realistic estimate, there are two contributions that should additionally be taken into account for the calculation of the uncertainty of the result of a fit process as in the present case. Three contributions are taken into account:

\begin{itemize}
\item the standard deviation of the mean based on a sample $s_A$ from the results of the four different measurements,
\item the fitting uncertainty $\Delta A_{\mathrm{Fitter}}$ returned by the program Fitter (see \citet{Zei22}), 
\item the uncertainty $\Delta A_{\mathrm{fixed}}$ of the magnetic dipole hfs constant $A$ which has been fixed during the fit:
\end{itemize}
\[
\Delta A_{\mathrm{exp}} = \sqrt{ (s_A)^2 + (\Delta A_{\mathrm{Fitter}})^2 + (\Delta A_{\mathrm{fixed}})^2} \,\, .
\]

If an $A_{\mathrm{exp}}$ value for a level is determined from two different lines, the weighted mean values $A_{\mathrm{mean}}$  of the results $A_{\mathrm{exp}}$  of the individual lines are calculated, using reciprocal squares of uncertainties as weights $w$:
\[
A_{\mathrm{mean}} = \frac{\sum_i{(A_{\mathrm{exp},\,i}\cdot w_i)}}{\sum_i{w_i}}
\]
with
\[
w_i = \frac{1}{(\Delta A_{\mathrm{exp},\,i})^2} \,\, .
\]

Here, again, the purely statistical approach does not reflect the uncertainty well. In order to take into account both the uncertainties of the individual values and the spread of the values when calculating the uncertainties, the following formula was used to calculate the uncertainties $\Delta A_{\mathrm{mean}}$ of the weighted mean values $A_{\mathrm{mean}}$ (see for example \citep{Rad69}: 
\[
\Delta A_{\mathrm{mean}} = \frac{\sqrt{\sum_i{(\Delta A_{\mathrm{exp},\,i}^2+(A_{\mathrm{exp},\,i}-A_{\mathrm{mean}})^2)\cdot w_i^2 }}}{\sum_i{w_i}} \,\, .
\]

Two lines, marked by an asterisk are so weak and noisy that the resultant $A$-values were not taken into account when calculating the mean values.

\begin{center}
\begin{table*}
\centering
\caption{New and revised magnetic dipole hyperfine structure constants $A$ for levels of odd and even parity of Tm~II. \label{tab:t3}}
\begin{tabular}{c r@{}l c r r r@{}l c r@{}l c c}
\hline \hline
&\multicolumn{2}{c}{$E$ in cm$^{-1}$} & $J$ &
\multicolumn{1}{c}{configuration} &
\multicolumn{1}{c}{term} &
\multicolumn{2}{c}{$\sigma_{\mathrm{\textcolor{blue}{obs}}}$ in cm$^{-1}$} &
\multicolumn{1}{c}{assignment} &
\multicolumn{2}{c}{$A_\mathrm{exp}$ in MHz} &
\multicolumn{2}{c}{$A_\mathrm{mean}$ in MHz} \\[1mm]
\hline
\multicolumn{11}{l}{\textbf{levels of odd parity}} \\                                                
&  \,  0&.00 &  4 &  
$4f$$^{13}$$(^{2}$$F_\mathrm{7/2}$)$6s_\mathrm{1/2}$ 
 & (7/2,1/2) & 26\,578&.78 &   \textcolor{violet}{A}    &   -1038&.3\,(1.6) &      &         \\[0.8mm]     
&  \,236&.95 & 3 & $4f$$^{13}$$(^{2}$$F_\mathrm{7/2}$)$6s_\mathrm{1/2}$
& (7/2,1/2) & 25\,743&.09 & \textcolor{blue}{C}      &     293&\,(6)     &      &         \\[0.8mm]     
& 8\,769&.68 & 2 & $4f$$^{13}$$(^{2}$$F_\mathrm{5/2}$)$6s_\mathrm{1/2}$
& (5/2,1/2) & 25\,628&.62 & \textcolor{orange}{B}    &     129&\,(4)     &      &         \\[0.8mm]     

\multicolumn{11}{l}{\textbf{levels of even parity}} \\                                                                                   
&18\,291&.37 & 4 & $4f$$^{12}$$(^{3}$$H_\mathrm{6}$)$5d6s$$(^{3}$$D_\mathrm{2}$)
& (6,2) & 18\,291&.38 & \textcolor{orange}{B}    &     -92&\,(4)     &      &         \\[0.8mm]     
&21\,608&.26 & 3 &$4f$$^{12}$$(^{3}$$H_\mathrm{6}$)$5d6s$$(^{3}$$D_\mathrm{3}$)
& (6,3) & 21\,608&.26 & \textcolor{orange}{B}    &     140&\,(3)     &      &         \\[0.8mm]     
&22\,308&.82 & 4 
&$4f$$^{12}$5$d$6$s$ &   &
22\,308&.84 & \textcolor{orange}{B}    &    -949&\,(10)    &      &         \\[-0.2mm]    
&       &     &     &   &    &
22\,071&.89 & \textcolor{teal}{D}      &    -948&\,(6)     &  -948\,(\sk{6}) &   \\[0.8mm]     
&23\,768&.84 & 5 &$4f$$^{12}$$(^{3}$$H_\mathrm{6}$)$5d6s$$(^{3}$$D_\mathrm{3}$)
& (6,3) & 23\,768&.85 & \textcolor{orange}{B}    &     -73&.8\,(2.1) &      &         \\[0.8mm]     
&23\,803&.28 & 4 
&$4f$$^{12}$5$d$6$s$ &   &
 23\,803&.29 & \textcolor{orange}{B}    &     -38&.6\,(2.0) &      &         \\[-0.2mm]    
&       &     &     &   &    &
23\,566&.34 & \textcolor{teal}{D}      &     -38&\,(7)     &   \,-38.6\,(\sk{1.9}) \\[0.8mm]     
&25\,257&.52 & 3  &$4f$$^{12}$$(^{3}$$F_\mathrm{4}$)$5d6s$$(^{3}$$D_\mathrm{1}$)
& (4,1)& 25\,257&.53 & \textcolor{orange}{B}    &    -682&.9\,(2.2) &      &         \\[-0.2mm]    
&       &     &     &   &    & 25\,020&.58 & \textcolor{teal}{D}      &    -681&\,(7)     &  -682.7\,(\sk{2.1})\\[0.8mm]     
&25\,980&.02 & 3 & $4f$$^{13}$6$p$ &   &
8\,355&.37 & \textcolor{violet}{A}    &    -656&\,(4)     &      &         \\[-0.2mm]    
&       &    &   &   &    & 
25\,980&.03 & \textcolor{orange}{B}    &    -653&\,(5)     &  -654\,(\sk{4})    \\[-0.2mm]    
&       &    &   &   &    & 
4\,266&.29 & \textcolor{violet}{A}*   &    -655&\,(20)*   &      &    
\\[0.8mm]     
&26\,574&.66 & 4 
&$4f$$^{12}$5$d$6$s$ &   &26\,574&.67 & \textcolor{orange}{B}    &    -206&.3\,(2.0) &      &         \\[-0.2mm]    
&       &    &   &  &  & 26\,337&.72 & \textcolor{teal}{D}      &    -207&\,(6)     &  -206.4\,(\sk{1.9})\\[0.8mm]     
&27\,009&.39 & 4 
&$4f$$^{12}$5$d$6$s$ &  &27\,009&.40 & \textcolor{orange}{B}    &    -812&\,(6)     &      &         \\[-0.2mm]    
&       &    &   &  &  &  26\,772&.46 & \textcolor{teal}{D}      &    -819&\,(6)     &  -816\,(\sk{5})    \\[0.8mm]     
&27\,254&.42 & 4 
&$4f$$^{12}$5$d$6$s$ & & 27\,017&.48 & \textcolor{teal}{D}      &  
-503&\,(6)     &      &         \\[0.8mm]     
&27\,702&.42 & 3 
&$4f$$^{12}$5$d$6$s$ & & 10\,077&.78 & \textcolor{violet}{A}    &    -711&\,(6)     &      &         \\[-0.2mm]    
&       &    &   &  & & 27\,702&.44 & \textcolor{orange}{B}    &    -703&\,(9)     &  -709\,(\sk{6})    \\[0.8mm]     
&28\,267&.88 & 3 
&$4f$$^{12}$5$d$6$s$ & & 28\,267&.88 & \textcolor{orange}{B}    &   
-698&\,(12)     & &              \\[-0.2mm]    
&       &    &   & & &28\,030&.94 & \textcolor{teal}{D}      &    -717&\,(10)    &  -709\,(\sk{10})   \\[0.8mm]     
&29\,183&.39 & 4 
&$4f$$^{12}$5$d$6$s$ & & 29\,183&.40 & \textcolor{orange}{B}    &    -479&\,(6)     &      &         \\[-0.2mm]    
&       &    &   & & &28\,946&.45 & \textcolor{teal}{D}      &    -484&\,(7)     &  -481\,(\sk{5})    \\[0.8mm]     
&29\,424&.98 & 3 
&$4f$$^{12}$5$d$6$s$ & &29\,424&.98 & \textcolor{orange}{B}    &    
-175&\,(6)     &      &         \\[0.8mm]     
&29\,967&.17 & 2 
&$4f$$^{12}$5$d$6$s$ & & 29\,730&.23 & \textcolor{teal}{D}      &   -1045&\,(8)     &      &         \\[0.8mm]     
&31\,036&.61 & 3 
&$4f$$^{12}$$(^{3}$$H_\mathrm{4}$)$5d6s$$(^{3}$$D_\mathrm{1}$)
& (4,1)& 22\,266&.94 & \textcolor{blue}{C}      &    -681&\,(11)    &      &         \\[0.8mm]     
&32\,500&.26 & 2 
&$4f$$^{12}$5$d$6$s$ & & 23\,542&.80 & \textcolor{violet}{A}    &    -735&\,(21)    &      &         \\[-0.2mm]    
&       &    &   & & &23\,730&.59 & \textcolor{blue}{C}*     &    -696&\,(54)*   &      &         \\[0.8mm]     
&33\,391&.55 & 4 
&$4f$$^{12}$5$d$6$s$ & & 24\,434&.09 & \textcolor{violet}{A}    &    -277&\,(18)    &      &         \\[0.8mm]     
&34\,398&.30 & 3 
&$4f$$^{12}$5$d$6$s$ & & 25\,440&.83 & \textcolor{violet}{A}    &    -676&\,(3)     &      &         \\[0.8mm]     
&35\,753&.72 & 3 
&$4f$$^{12}$5$d$6$s$ & & 26\,984&.03 & \textcolor{blue}{C}      &    -652&\,(8)     &      &         \\[0.8mm]     
&36\,041&.02 & 3 
&$4f$$^{12}$5$d$6$s$ & & 27\,271&.34 & \textcolor{blue}{C}      &    -992&\,(7)     &      &         \\[0.8mm]     
&36\,132&.08 & 3 
&$4f$$^{12}$5$d$6$s$ & & 27\,174&.59 & \textcolor{violet}{A}    &    -391&\,(7)     &      &         \\[0.8mm]     
&36\,394&.64 & 2 
&$4f$$^{12}$5$d$6$s$ & & 27\,437&.16 & \textcolor{violet}{A}    &    -358&\,(4)     &      &         \\[0.8mm]     
&36\,868&.99 & 3
&$4f$$^{12}$5$d$6$s$ & & 28\,099&.31 & \textcolor{blue}{C}      &   -1026&\,(5)     &      &         \\[0.8mm]     
&38\,093&.53 & 4 
&$4f$$^{12}$5$d$6$s$ & & 29\,136&.07 & \textcolor{violet}{A}    &    -584&\,(11)    &      &         \\[0.8mm]     
&38\,361&.24 & 4 
&$4f$$^{12}$5$d$6$s$ & & 29\,403&.78 & \textcolor{violet}{A}    &    -595&\,(4)     &      &         \\            
\hline
\end{tabular}
\tablecomments{configuration and term in $JJ$ coupling according to \citep{Mar78};
assignment: line group category,} see text for explanation; *: weak and noisy lines not taken into account when calculating the mean values of $A$ constants.
\end{table*}

\end{center}

\section{Reflection on the revised hfs $A$ constants} \label{sec:reflections}

\begin{figure}
\includegraphics[scale=0.45]{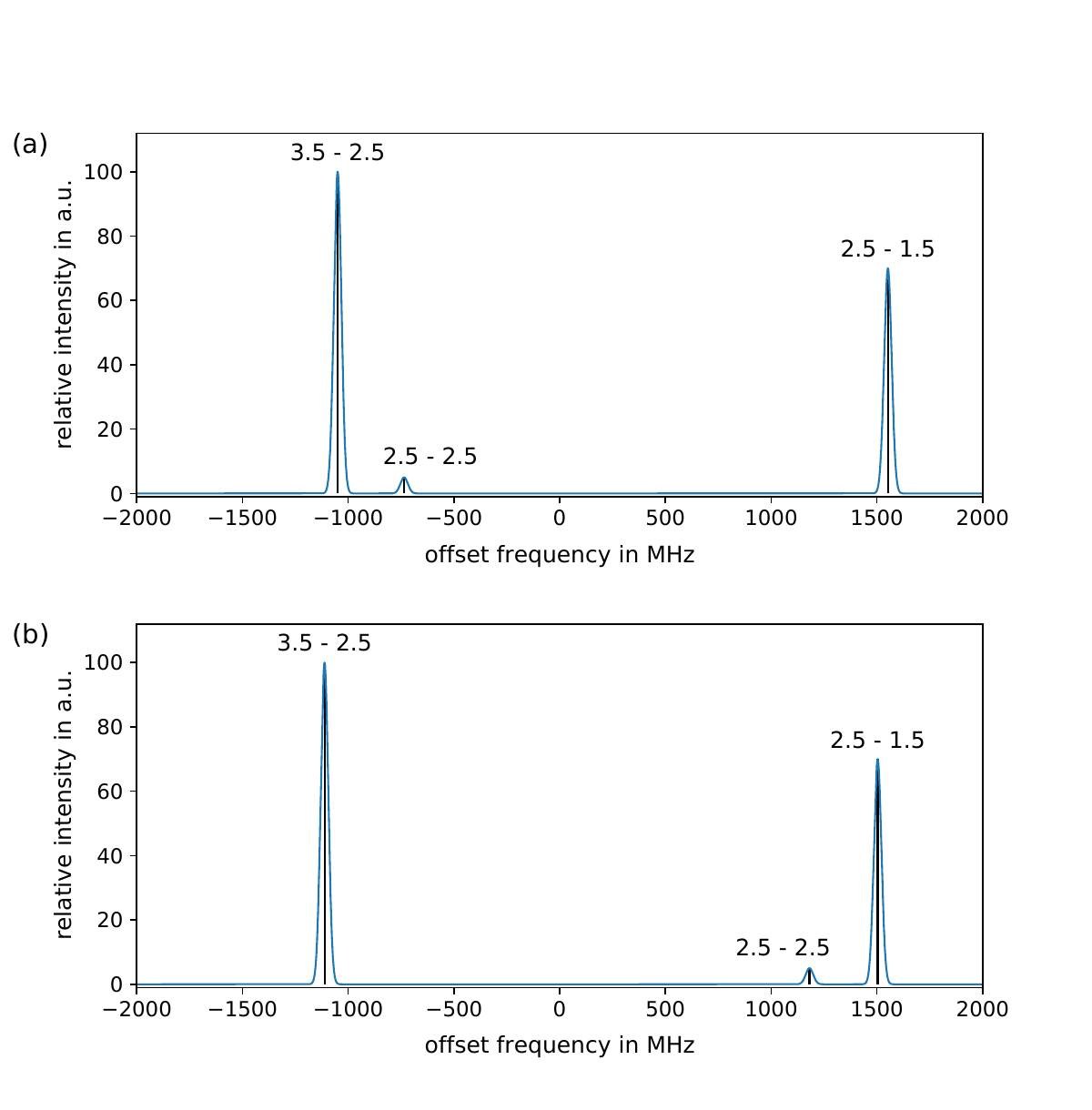}
\caption{Simulation of the transition 8769.68~cm$^{-1}$ ($J = 2$) to 25980.02~cm$^{-1}$ ($J = 3$) at $\sigma_{\mathrm{Ritz}} = 17\,210.34$~cm$^{-1}$ or $\lambda_{\mathrm{air}} = 580.885$~nm, respectively, using a Gaussian Profile and a FWHM of 40~MHz.\\
a) simulated with our new hfs constants $A$,\\
b) simulated with the hfs constants $A$ from \citet{Man89}. \\		
\label{fig:f3}}
\end{figure}

As mentioned above, we have declared two values from the paper of \citet{Man89} to be incorrect.  
This concerns the two levels at 8\,769.68~cm$^{-1}$ ($A$=914.9(1.0)MHz) and 25\,980.02~cm$^{-1}$ ($A$=-89.8(0.7)MHz).
We were very concerned with the question of how it is possible that such a high precision measurement method as collinear fast-ion-beam laser
spectroscopy used in the work of \citet{Man89} can produce incorrect values for the hfs constants.
To explore this question, we first considered what transitions might have been studied in \citet{Man89}. The paper mentions a dye laser but does not give a list of the lines examined. The wavelengths or wave numbers mentioned in the figure captions suggest that the dye R6G was used. 
Using the Ritz combination principle, we have calculated all possible transitions based on the levels mentioned in \citet{Man89} in the wavelength range of this dye. 
The following three transitions were found involving the levels with the incorrect $A$-values:
\begin{enumerate}
\item[(\textbf{a})] 8769.68~cm$^{-1}$ to 25980.02~cm$^{-1}$\\ at $\sigma_{\mathrm{Ritz}} = 17210.34$~cm$^{-1}$ or $\lambda_{\mathrm{air}} = 580.885$~nm;
\item[(\textbf{b})]  8957.47~cm$^{-1}$ to 25980.02~cm$^{-1}$\\ at $\sigma_{\mathrm{Ritz}} = 17022.55$~cm$^{-1}$ or $\lambda_{\mathrm{air}} = 587.293$~nm;
\item[(\textbf{c})] 8769.68~cm$^{-1}$ to 26578.77~cm$^{-1}$\\ at $\sigma_{\mathrm{Ritz}} = 17809.09$~cm$^{-1}$ or $\lambda_{\mathrm{air}} = 561.355$~nm.
\end{enumerate} 
These three lines are not visible in our spectrum. In the case of (\textbf{b}) and (\textbf{c}), only one of the levels with the incorrect $A$ value is involved and we have no explanation at all for a possible deviation. 
For the transition in case (\textbf{a}), the $A$ of both levels are declared to be incorrect. 
For this transition, we simulated the hfs with our new $A$ values and with the $A$ values of \citet{Man89} using a Gaussian profile and a FWHM that corresponds to the FWHM of the curves in the figures in the paper of \citet{Man89}. These simulations are shown in Figure \ref{fig:f3}. 
The offset frequency is given relative to the centre of gravity of the hfs.
Looking at these two figures, one can see that the total splitting is the same in both cases. The distance between the weak component and the closest strong component is also the same, except for the fact that in the first case the distance to the strongest component is measured and in the second case the distance to the second strongest component is measured. 
We venture here the bold assertion that in the work of \citet{Man89} the transition (\textbf{a}) was used to determine the $A$ constants of the two levels 8\,769.68~cm$^{-1}$ and 25\,980.02~cm$^{-1}$ and that the distances from the weak component to the two strong components were inadvertently reversed. 

Based on this assumption, we can now calculate the distances of the hfs components with the $A$-values from \citep{Man89}, swap the allocation of the distances between the components and re-calculate the corrected $A$-values. 
The resulting values are $A = 125.7$~MHz for the level 8\,769.68~cm$^{-1}$ and $A = -653.5$~MHz for the level 25\,980.02~cm$^{-1}$, which agree with our values within the error limits. 
\medskip

In the paper of \citet{Man89} theoretical values from multiconfiguration Dirac-Fock (MCDF) ab initio calculations are given for four lower-lying levels. The following statement was made in \citep{Man89} about these theoretical values: "... hfs of configurations explicitly containing open s-shells are in substantial disagreement with MCDF calculations."
The deviation was particularly evident for the level 8769.68~cm$^{-1}$ whose $A$ constant we have corrected here. 
Our new value of $A_{\mathrm{exp}} = 129\,(4)$~MHz is much closer to the theoretical value of $A_{\mathrm{MCDF}} = -67$~MHz than the previous value of $A_{\mathrm{exp}} = 914.9\,(1.0)$~MHz from \citet{Man89}. Thus, with our new $A_{\mathrm{exp}}$ value, the statement from \citet{Man89} 
quoted above concerning the MCDF calculations is
no longer true.
\\

\section{Conclusion and outlook}

The hfs of 40 selected Tm~II lines in high-resolution FT spectra have been investigated. All except three of the selected lines show
two clearly separated strong hfs components. Based on these 40 lines and using the known magnetic dipole hfs constant $A$ from the literature \citep{Man89}, hfs constants $A$ for 29 energy levels were determined. 
Of these 29 hfs $A$ values, 27 were measured for the first time. The remaining two $A$ constants, which were previously measured by \citet{Man89} were found to be incorrect. The revised values for these have been provided.   
Among the levels with newly determined $A$ values are the ground state and the second level of the ground term $4f^{13}(^2F_{7/2})\,6s_{1/2}$, both of which are involved in many strong transitions and therefore of particular interest for astrophysical studies.

Based on the here presented set of hfs constants a comprehensive analysis of Tm spectral lines is planned. It is intended to carry out an extensive weighted least-squares fit of experimental peak wavenumbers of strong hfs components for analysis of fine structure and hyperfine structure of atomic and singly ionized Tm.

With our new and revised values, the hfs constants $A$ are now known for all four levels of the ground configuration $4f^{13}\,6s$. 
It would be interesting to carry out new semi-empirical or ab initio calculations and compare the results with our new experimental values. It should be noted that the hfs constants determined for the transitions of Tm\,{\sc ii} in this study can be used to generate complete line component patterns for these transitions. This will greatly assist stellar spectroscopists in incorporating the latest data into their spectral synthesis programs to obtain accurate element abundances for Tm\,{\sc ii}.

\bigskip

\begin{acknowledgments}
Acknowledgments:\\
This work has been supported by the Istanbul University Scientific Research Project via Project No: 33657 and also funded by the Scientific Research Projects Coordination Unit of Istanbul University Project No: 30048, as well as by Latvian Council of Science, project No. lzp-2020/1-0088:``Advanced spectroscopic methods and tools for the study of evolved stars''.
We thank Chris Sneden for providing the Keck spectra used to construct Figure 1 in the paper.
\end{acknowledgments}


\end{document}